\documentclass[9pt,twocolumn,twoside]{optica}
\setboolean{shortarticle}{true}
\setboolean{minireview}{false}
\usepackage{braket}
\usepackage{color,soul}

\title{Persistent energy-time entanglement covering multiple resonances of an on-chip biphoton frequency comb}

\author[1,2,4,†,*]{Jose A. Jaramillo-Villegas}
\author[1,2,†]{Poolad Imany}
\author[1,2]{Ogaga D. Odele}
\author[1,2]{Daniel E. Leaird}
\author[5]{Zhe-Yu Ou}
\author[1,3]{Minghao Qi}
\author[1,2,3]{Andrew M. Weiner}

\affil[1]{School of Electrical and Computer Engineering, Purdue University, West Lafayette, Indiana, USA}
\affil[2]{Purdue Quantum Center, Purdue University, West Lafayette, Indiana, USA}
\affil[3]{Birck Nanotechnology Center, Purdue University, West Lafayette, Indiana, USA}
\affil[4]{Facultad de Ingenierías, Universidad Tecnológica de Pereira, Pereira, Risaralda, Colombia}
\affil[5]{Department of Physics, Indiana University-Purdue University Indianapolis, Indianapolis, Indiana, USA}
\affil[*]{Corresponding author: jjv@purdue.edu}

\ociscodes{(270.5585) Quantum information and processing; (130.3990) Micro-optical devices; (140.4780) Optical resonators.}

\begin{abstract}
We investigate the time-frequency signatures of an on-chip biphoton frequency comb (BFC) generated from a silicon nitride microring resonator. Using a Franson interferometer, we examine, for the first time, the multifrequency nature of the photon pair source in a time entanglement measurement scheme; having multiple frequency modes from the BFC results in a modulation of the interference pattern. This measurement together with a Schmidt mode decomposition shows that the generated continuous variable energy-time entangled state spans multiple pair-wise modes. Additionally, we demonstrate nonlocal dispersion cancellation, a foundational concept in time-energy entanglement, suggesting the potential of the chip-scale BFC for large-alphabet quantum key distribution.
\end{abstract}

\setboolean{displaycopyright}{true}

\begin{document}

\maketitle


Quantum information processing (QIP) promises to improve the security of our communications as well as to solve some algorithms with exponential complexity in polynomial time \cite{nielsen2010quantum}. The fundamental unit of quantum information is based on a superposition of two states, the so-called qubit. It has been proposed to extend this concept to a superposition of many states (known as a high-dimensional state) for higher density information encodings, fault-tolerant quantum computing, and even for secure protocols in dense quantum key distribution.

Entangled photon pairs have been demonstrated as one of the most promising platforms for implementing QIP systems. In recent years, generation of entangled photons on chip has gained attention because of its reduced cost and compatibility with semiconductor foundries \cite{sharping2006generation,clemmen2009continuous,ramelow2015silicon, grassani2015micrometer}. Despite Biphoton Frequency Combs (BFC) have been generated using cavity-filtered broadband biphotons \cite{lu2003mode} and cavity-enhanced spontaneous parametric down conversion \cite{scholz2007narrow}, it was only recently that on-chip microresonators have been used to generate entangled photons in the form of a BFC \cite{reimer2014integrated,reimer2016generation,mazeas2016high}. These demonstrations suggest the use of chip-scale sources for high-dimensional quantum processing \cite{xie2015harnessing,bessire2014versatile,lukens2016frequency}. However, studies such as \cite{ramelow2015silicon,reimer2016generation,mazeas2016high} only focused on single sideband pairs---the multifrequency nature of their sources (important for high-dimensional quantum processing) was not explored. In this Letter, we present the first examination of the time-frequency signatures of an on-chip BFC generated from a silicon nitride microring resonator. Through Franson interferometry and a demonstration of nonlocal dispersion compensation, we are able to examine the multifrequency nature of our photon-pair source. 

\begin{figure}[!t]
\centering
\includegraphics[width=\linewidth]{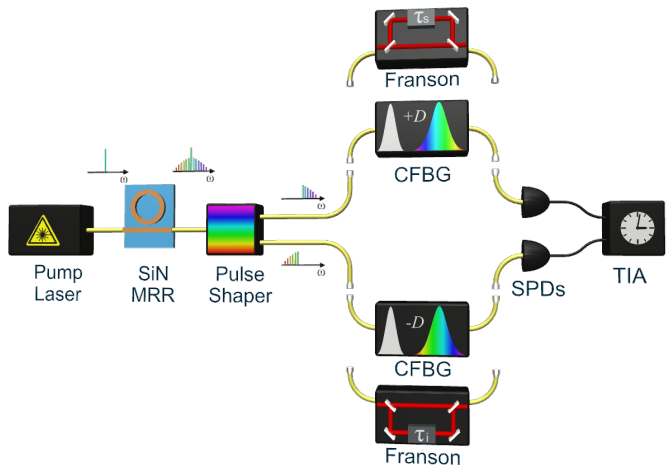}
\caption{Experimental setup. The biphoton frequency comb is generated by seeding a silicon nitride microring mesonator (SiN MRR) with a continuous wave tunable laser. A pulse shaper splits signal and idler sidebands. Chirped Fiber Bragg Gratings (CFBG) with opposite dispersions are used for the nonlocal dispersion cancellation experiment; unbalanced (“Franson”) interferometers are used to measure two-photon interference signatures. Detection is performed using Single Photon Detectors (SPD) and a Time Interval Analyzer (TIA).}
\label{fig:ExperimentalSetup}
\end{figure}

\begin{figure}[!ht]
\centering
\includegraphics[width=0.85\linewidth]{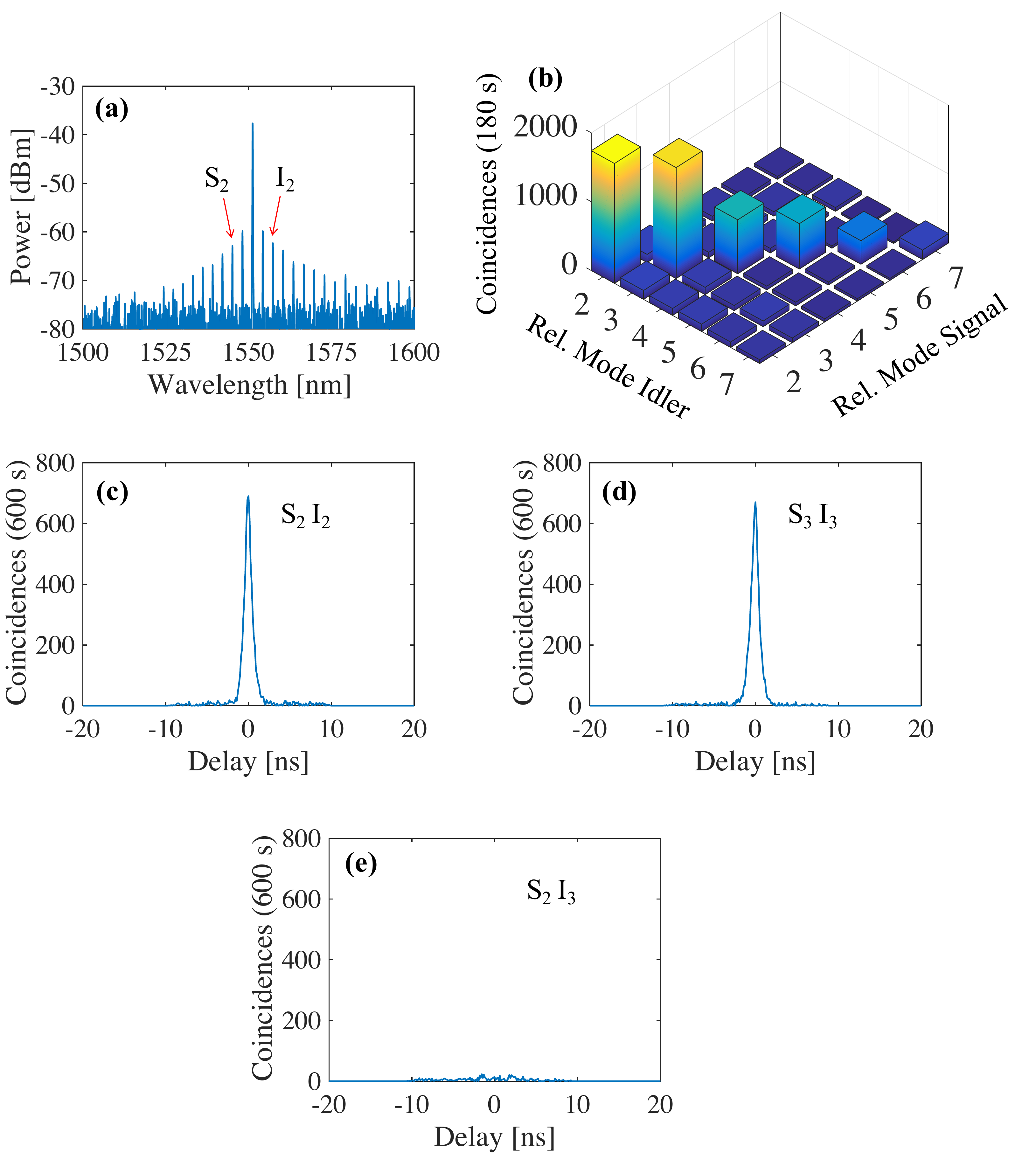}
\caption{Characterization of biphoton frequency comb. (a) Optical spectrum. (b) Joint spectral intensity between signal photons with relative modes from 2 to 7, $S_{2-7}$, and idler photons with relative modes from 2 to 7, $I_{2-7}$. Time correlations for (c) $S_2I_2$, (d) $S_3I_3$ and (e) $S_2I_3$.}
\label{fig:JointSpectralIntensity}
\end{figure}

Figure \ref{fig:ExperimentalSetup} is a depiction of our experimental setup. To generate our BFC, we use a tunable continuous-wave (CW) laser to pump a microring at the resonance located at the frequency $\omega_{\textrm{\fontsize{6}{0}\selectfont{p}}}$ ($\lambda_{\textrm{\fontsize{6}{0}\selectfont{p}}} \sim$ 1550.9 nm). Through spontaneous four-wave mixing process, two pump photons decay into one photon at higher frequency called signal, and another photon at a lower frequency called idler. Due to the resonant structure of the microring, this process occurs within narrow cavity resonances, described by a lineshape function $\Phi(\Omega)$ and the separation between these resonances is $\Delta\omega$, the free spectral range (FSR) of the microring. As a consequence of energy conservation, the signal and idler photons are highly correlated in frequency and time. This results in a comb-like photon pair spectrum or BFC as shown in Fig. \ref{fig:JointSpectralIntensity}a. In this figure, we observe a roll-off in the power of the sidebands as we move away from the pump frequency. This is due to the finite dispersion of the microring, which reduces the overlap of the energy-matched resonances as we move away from the central frequency (see Supplementary Information S.2). As a consequence of energy conservation, we expect that the quantum state of the generated biphotons may be written as:

\begin{equation}
\Ket{\Psi} = \sum_{k=1}^{N} { \alpha_{k} \ket{k,k}_{\textrm{\fontsize{6}{0}\selectfont{SI}}} }
\label{eq:State}
\end{equation}
where
\begin{equation}
\ket{k,k}_{\textrm{\fontsize{6}{0}\selectfont{SI}}} = \int{d\Omega  \hspace{2pt} \Phi(\Omega - k\Delta\omega) \Ket{\omega_{\textrm{\fontsize{6}{0}\selectfont{p}}}+\Omega,\omega_{\textrm{\fontsize{6}{0}\selectfont{p}}}-\Omega }_{\textrm{\fontsize{6}{0}\selectfont{SI}}}}
\label{eq:withinstate}
\end{equation}

\noindent which represents a superposition of many pair-wise signal and idler frequency combinations symmetric around the pump frequency. While the complex factors $\alpha_k$ give the amplitude weighting and phase of the various signal-idler sideband pairs, the experiments we conduct are insensitive to phase coherence between multiple sidebands, and thus depend only on $|\alpha_k|$. The pulse shaper \cite{weiner2000femtosecond} in Fig. \ref{fig:ExperimentalSetup} is used as a programmable frequency filter to separate the desired signal and idler photons and to route them to a pair of single-photon detectors. Additional band stop filters (not shown in Fig. \ref{fig:ExperimentalSetup}) are used at the output of the microring to remove the pump which also eliminates the first sideband pair, and as such, we are only able to examine from sideband pairs 2 and beyond. 


To verify the correlations of our comb-like photon pairs, we first select the signal and idler photons of the 2\textsuperscript{nd} sideband pair ($S_2I_2$). Using the pair of single-photon detectors along with an event timer, we record the relative arrival time between signal and idler photons as coincidences. Some accidental events are also registered as a result of detecting background (uncorrelated) photons---any event that is not due to a signal and idler from the same photon-pair is considered an accidental (dark counts, signal and idler from different pairs, or any such combination). Figure \ref{fig:JointSpectralIntensity}c shows the measured coincidences (accidentals were not subtracted); the sharp peak with a full width at half maximum (FWHM) of $\sim600$ ps corresponds to the temporal correlation. This is in agreement with the expected correlation time which is calculated from the inverse of the resonance linewidth ($\sim2\pi\times270$ MHz). Furthermore, to show correlations exist only between energy-matched frequencies, we measured the coincidences between the 3\textsuperscript{rd} signal and 3\textsuperscript{rd} idler ($S_3I_3$) (Fig. \ref{fig:JointSpectralIntensity}d), and the coincidences between the 2\textsuperscript{nd} signal and 3\textsuperscript{rd} idler ($S_2I_3$) (Fig. \ref{fig:JointSpectralIntensity}e). The absence of a coincidence peak between the 2\textsuperscript{nd} signal and 3\textsuperscript{rd} idler reveals a lack of correlation between mismatched frequencies. In addition, we were able to obtain a high coincidence to accidental ratio (CAR) of 52 for the 3\textsuperscript{rd} sideband pair, without compensating for the losses in our setup; if we take into account the losses from the biphoton generation stage up till detection, the corrected CAR would be 655.

In order to show the spectro-temporal correlations across the photon-pair spectrum, we measured the Joint Spectral Intensity (JSI) by using the pulse shaper as a programmable frequency filter to route different sidebands to the pair of detectors. The time correlation measurement was repeated between all combinations of the sideband pairs from 2 to 7 ($S_{2-7}I_{2-7}$). Figure \ref{fig:JointSpectralIntensity}b shows the measured JSI, which provides a strong confirmation that time correlations only appear in the energy matched sidebands. 


\begin{figure}[!ht]
\centering
\includegraphics[width=0.9\linewidth]{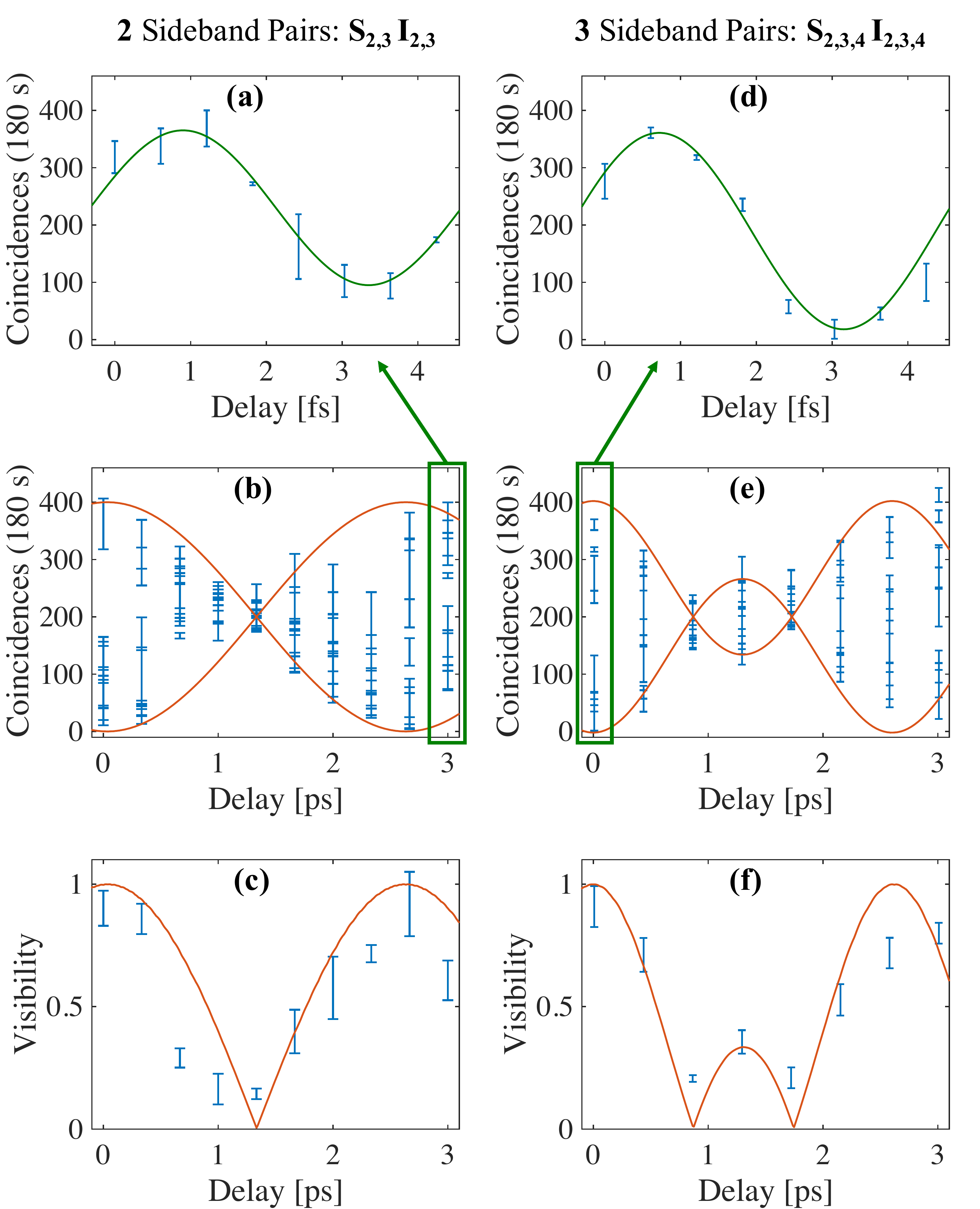}
\caption{Two-photon interference experiment of biphoton frequency comb. Interference fringes measured with a Franson interferometer for (a) small and (b) large $\tau_\textrm{d}$ ranges with two sideband pairs. (c) The visibility of the sinusoidal fringes in (b). Registered coincidences using three sideband pairs for (d) small and (e) large $\tau_\textrm{d}$ ranges. (f) The visibility of the sinusoidal fringes in (e). The blue error bars are the experimental results, the red curves in (b), (c), (e) and (f) are theoretical predictions, and the green curves in (a) and (d) are sinusoids fitted to the experimental data.}
\label{fig:FransonInterferometer}
\end{figure}

Next, we examine the signatures of using multiple frequency bins of the BFC in a two-photon interference experiment \cite{franson1989bell}. In this experiment, a Franson interferometer—two unbalanced Mach-Zehnder interferometers (MZI)—are placed in the signal and idler paths (Fig. \ref{fig:ExperimentalSetup}). The relative delays between long and short arms in the signal and idler interferometers, defined as $\tau_\textrm{s}$ and $\tau_\textrm{i}$, respectively, are approximately 6 ns. This value is much less than the coherence time of the pump ($\sim 1$ $\mu$s) but greater than the coherence time of a single photon ($\sim 1$ ns) to avoid self-interference. Here, the difference between $\tau_\textrm{s}$ and $\tau_\textrm{i}$ is defined as $\tau_\textrm{d}$. When $\tau_\textrm{d}$ is near to 0, the arrival time difference between signal and idler photons traveling through the long arms is approximately the same as when they both travel through the short arms. In consequence, we have path indistinguishability in this detection scheme using gated detection to only register the mentioned events ($\Ket{SS}$ and $\Ket{LL}$), discarding the events in which the signal photon travels the long path and the idler photon travels the short path and vice versa ($\Ket{SL}$ and $\Ket{LS}$). By setting $\tau_\textrm{d}$ to 0 and varying $\tau_\textrm{s}$ and $\tau_\textrm{i}$ together, we retrieve a sinusoidal coincidence pattern with a period of  $\sim 2.5$ fs (half the period of the pump laser) and a visibility of $80\% \pm 5\%$ for all signal and idler sidebands (see Supplementary Information S.3).  (Here, instead of using two interferometers, we sent signal and idler sidebands ($S_{2-4}I_{2-4}$) through one interferometer and they were split at the output using a pulse shaper.) The interference pattern is similar to those reported in previous microcavity BFC experiments that examined only a single signal-idler pair \cite{ramelow2015silicon, reimer2016generation,mazeas2016high}. The high visibility \cite{thew2004bell} shows for the first time that we maintain energy-time entanglement even for biphotons consisting of a multiplicity of sideband mode pairs. 

To see the multifrequency signature,  we fix $\tau_\textrm{s}$ and vary $\tau_\textrm{i}$ (hence varying $\tau_\textrm{d}$) in a two interferometer experiment. Sweeping $\tau_\textrm{d}$ over a small range results in a sinusoidal interference pattern in the registered coincidences, as shown in Figs. \ref{fig:FransonInterferometer}a,d.  The period of $\sim5$ fs corresponds to the average optical carrier frequency. For Figs. \ref{fig:FransonInterferometer}a-c we use two pairs of sidebands ($S_{2-3}I_{2-3}$). By varying $\tau_\textrm{d}$ over a larger range, we observe modulation and revival of the envelope of the two-photon fringes with a period of 2.6 ps, corresponding to the inverse of the microring’s FSR. In contrast, the fringe envelope for an individual signal-idler pair should decay smoothly on the sub-ns time scale of Fig. \ref{fig:JointSpectralIntensity} \cite{ramelow2015silicon,reimer2016generation}; the observation of picosecond scale modulation arises from the superimposed contributions of multiple signal-idler frequency bins. For two signals and idlers, the coincidence rate, $C$, is of the form
\begin{equation}
C \propto 1+\cos\left(2\omega_0\tau_\textrm{s} + \frac{\omega_{i_2}+\omega_{i_3}}{2}\tau_\textrm{d}\right)\cos\left( \frac{\omega_{i_2}-\omega_{i_3}}{2}\tau_\textrm{d}\right)
\label{eq:Coincidences}
\end{equation}
where $\omega_0$ is the pump frequency and $\omega_{i_n}$ is the idler frequency of the $n$-th sideband pair. The envelope and visibility of the coincidences agree with the quantum mechanical model, eq. \ref{eq:Coincidences}. We repeated this experiment for 3 sideband pairs ($S_{2-4}I_{2-4}$)---the intensities of these sidebands were equalized with the pulse shaper in order to have almost equal contribution in the interference experiment. With increased number of sideband pairs, the visibility vs. delay curve becomes sharper (Fig. \ref{fig:FransonInterferometer}d-f), again in agreement with theory (see Supplementary Information S.3). The visibility of the fringes in Fig. \ref{fig:FransonInterferometer} is as high as 92\% $\pm$ 13\%. The modulation in the fringe envelope, down to 14\% $\pm$ 2\%, gives us clear evidence of equal contribution of these sidebands in the coincidence pattern.

After examining pairwise time-entanglement, we calculate the Schmidt number based on the measured JSI. The degree of correlations, calculated via Schmidt decomposition, gives us a Schmidt number lower-bound ($K_{\textrm{min}}$) of 4.0 \cite{eckstein2014high}. Therefore, we are able to corroborate a high frequency correlation of our photons, since $K_{\textrm{min}}$ is greater than one. If the off-diagonal terms are all set to zero in our measured JSI, we obtain a Schmidt number of 4.08, which confirms that we have a very low number of accidentals in our measurements. This Schmidt number can also be interpreted as the number of transferable bits ($\log_2(K_{\textrm{min}})=2$) with a photon pair. We note that while we have large continuous variable energy-time entanglement under pairs of resonance modes as suggested by $\Phi(\Omega)$ in Eq. \ref{eq:withinstate}---which would result in a very high Schmidt number---our measurement technique does not allow access to the fine structure under each resonance. Thus, the Schmidt analysis is done on $k$ but time-energy entanglement is on $\ket{k,k}_{\textrm{\fontsize{6}{0}\selectfont{SI}}}$. Therefore, our Schmidt number bound indicates that the continuous variable energy-time entangled state spans 4 effective modes (given by $\alpha_k$).


\begin{figure}[!ht]
\centering
\includegraphics[width=0.9\linewidth]{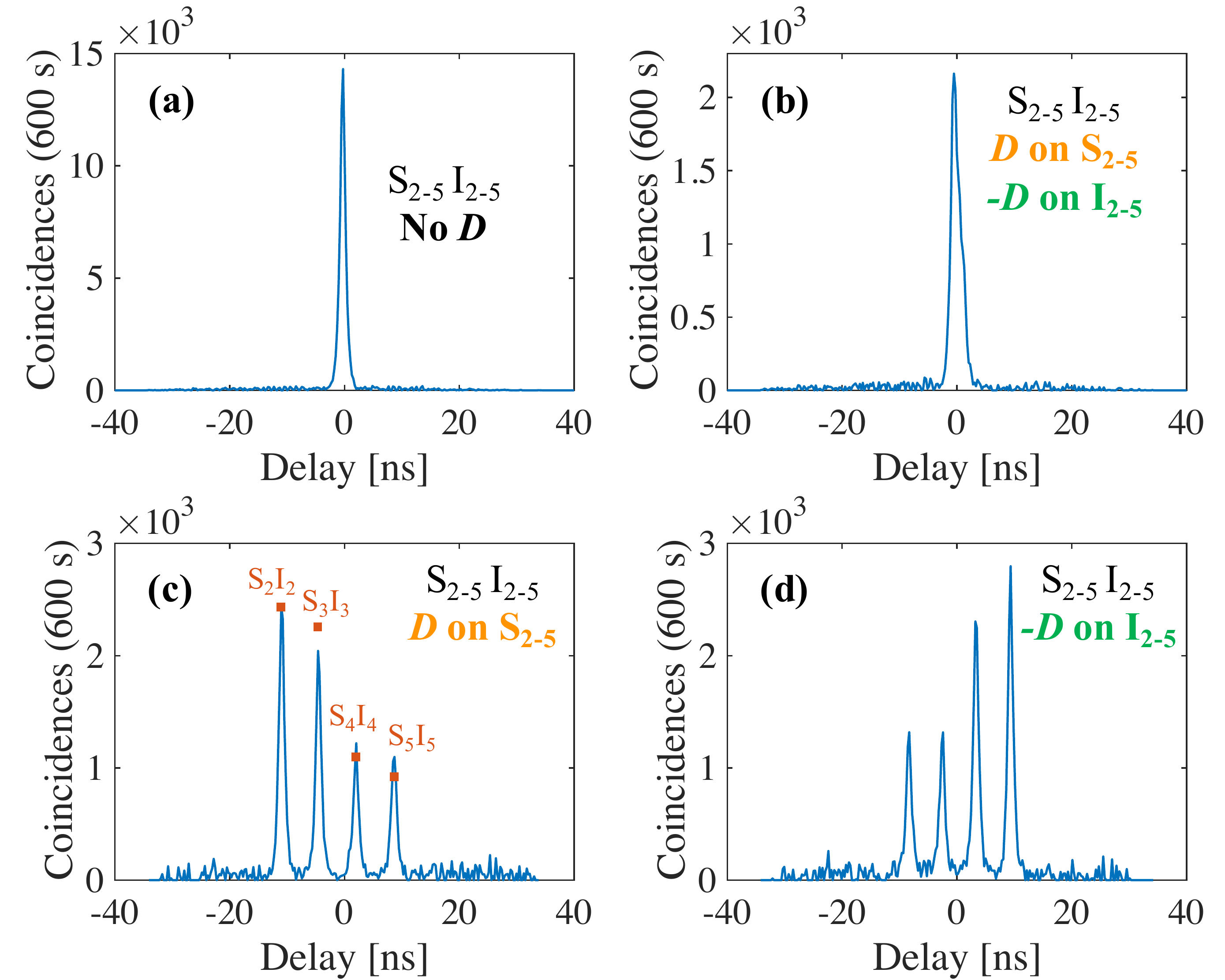}
\caption{Nonlocal dispersion cancellation experiment. Time correlation of four sideband pairs $S_{2-5}I_{2-5}$ (a) without dispersion, (c) with positive dispersion of $D=2$ ns/nm applied on signal photons $S_{2-5}$, (d) with negative dispersion of $D=-2$ ns/nm applied on idler photons $I_{2-5}$, and (b) with both dispersions applied at the same time as shown in Fig. \ref{fig:ExperimentalSetup} for nonlocal dispersion cancellation. In these plots, accidentals were subtracted and the effect of a finite detection-gate width (which results in a roll-off in coincidences as one moves away from zero delay) was compensated. The red squares in (b) represent the diagonal of the JSI in Fig \ref{fig:JointSpectralIntensity}b for the sideband pairs $S_{2-5}I_{2-5}$, normalized to the maximum of the blue plot for ease of visualization.}
\label{fig:NonlocalCancellationDispersion}
\end{figure}

We proceed to examine the potential of the BFC for quantum key distribution by demonstrating a nonlocal dispersion cancellation measurement \cite{franson1992nonlocal}, wherein the correlation peak maintains its undispersed form even though the signal and idler photons are dispersed. This nonlocal dispersion cancellation effect can enhance security in QKD by serving as a non-orthogonal basis to direct time-correlation measurements \cite{mower2013high,lee2014entanglement}. First, we use four sideband pairs ($S_{2-5}I_{2-5}$) and measure the correlation function in the absence of dispersion (Fig. \ref{fig:NonlocalCancellationDispersion}a). Next, we apply dispersion of 2 ns/nm (using a chirped fiber Bragg grating, which provides dispersion equivalent to that of $\sim$ 120 km of standard single mode fiber, but with a loss of only 3 dB) to only the signal sidebands of the BFC; this results in a measurement of four correlation peaks corresponding to the four different sideband pairs. The peaks are spaced by 6 ns, an expected outcome since different frequencies travel at different speeds in a dispersive medium (Fig. \ref{fig:NonlocalCancellationDispersion}c). Applying the opposite dispersion to only the idler sidebands will result in a similar outcome but with opposite sign of delay variation (Fig \ref{fig:NonlocalCancellationDispersion}d). These measurements with the separated correlation peaks are equivalent to a frequency-to-time mapping of our BFC, thus enabling us to resolve the JSI in the temporal basis \cite{valencia2002entangled}. To emphasize this equivalence, the diagonal terms of the JSI for sideband pairs ($S_{2-5},I_{2-5}$) are normalized to the maximum of Fig. \ref{fig:NonlocalCancellationDispersion}c and plotted as red squares; the good agreement with the correlation peaks in time provides a quantitative confirmation of frequency-to-time mapping. When we apply both dispersive media (positive dispersion on the signals and negative dispersion on the idlers), we expect nonlocal cancellation of the dispersion. As shown in Fig. \ref{fig:NonlocalCancellationDispersion}b, this behavior is clearly observed:  the coincidence plot collapses back into a single peak, with an improvement in the peak-to-background ratio evident despite the extra loss incurred through the introduction of a second CFBG.


In conclusion, we explore the time and frequency signatures of a biphoton frequency comb generated in a silicon nitride microring resonator. Two photon interference experiments demonstrate that continuous variable energy-time entanglement persists for biphotons spanning a multiplicity of discrete frequency modes and reveal interferometry signatures characteristic of the multiple frequency mode content. A Schmidt number analysis indicates that the continuous variable energy-time entangled state spans four effective discrete frequency modes. This suggests potential utility to high-dimensional frequency-bin coding recently proposed for quantum information processing \cite{lukens2016frequency}. Higher dimensionality should be possible through dispersion engineering, which would result in a BFC with increased number of frequency bins with a more homogeneous flux of photons in a larger bandwidth.

\paragraph{Funding.} National Science Foundation (ECCS-1407620). JAJ acknowledges support by Colciencias and Fulbright Colombia.

\paragraph{Acknowledgment.} We acknowledge Xiaoxiao Xue and Yi Xuan for providing the microring sample and thank Keith McKinzie, Nathan O’Malley, and Joseph M. Lukens for comments and discussions. $^\dagger$These authors contributed equally to this work.

\noindent See \href{link}{Supplement 1} for supporting content.


\bibliography{BFC}

\bibliographyfullrefs{BFC}


\end{document}